\documentclass[preprint,
secnumarabic,amssymb, nobibnotes, aps, pre,superscriptaddress,
]{revtex4-2}
\usepackage{soul,color,xcolor}

\usepackage{amsmath,amsthm}
\usepackage{array}
\usepackage{subfigure}
\usepackage{graphicx}
\usepackage{caption}
\usepackage{bm}
\usepackage[colorlinks,
          linkcolor=black,
            citecolor=black,
            urlcolor=blue
           ]{hyperref}
\usepackage{comment}
\usepackage{mathrsfs}
\usepackage{physics}
\usepackage{xcolor} % For changing text color
\usepackage{etoolbox}

\usepackage{multirow}
\usepackage{booktabs}
\usepackage{siunitx}

\begin{document}

%\captionsetup[figure]{labelfont={bf},name={Figure},labelsep=period}
\captionsetup{justification=raggedright}
\title{Emergence of Periodic Potential for Point Defects in a 2D Hexagonal Colloidal Lattice}

\author{Xicheng Huang}
\affiliation{Department of Physics, College of Sciences, Shanghai University, Shanghai 200444, China}
\affiliation{College of Biomedical Engineering, Sichuan University, Chengdu 610065, China}
\author{Zefei Liu}
\affiliation{College of Mechanical Engineering, Beijing Institute of Technology, Zhuhai 519088, China}
\author{Yong-Cong Chen}
\email[Correspondence: ]{chenyongcong@shu.edu.cn}
\affiliation{Shenzhen International Quantum Academy (IQASZ), Shenzhen 518048, China}
\affiliation{Department of Physics, College of Sciences, Shanghai University, Shanghai 200444, China}
\affiliation{College of Biomedical Engineering, Sichuan University, Chengdu 610065, China}
\author{Guohong Yang}
\affiliation{Department of Physics, College of Sciences, Shanghai University, Shanghai 200444, China}
\author{Ping Ao}
\email[Correspondence: ]{aoping@scu.edu.cn}
\affiliation{College of Biomedical Engineering, Sichuan University, Chengdu 610065, China}

\begin{abstract}
We examined the Brownian motion of point defects in a two‑dimensional hexagonal colloidal crystal, going beyond the conventional treatment that assumes constant diffusion coefficients. By extracting the spatially varying drift vector and diffusion matrix directly from experimental trajectories, we uncovered richer behavior than predicted by the simple diffusive limit. Within a general stochastic‑dynamics framework, these measurements revealed an effective stochastic potential landscape shaped by the crystal’s periodic structure. The energy differences between its local minima were consistent, to within an order of magnitude, with previous experimental estimates. Simulations of stochastic trajectories on this reconstructed landscape reproduced the essential features of the observed defect motion. This study illustrates how combining time‑series extraction with theoretical analysis can expose effective energy landscapes and provide a powerful route to understanding complex dynamics in colloidal systems.
\end{abstract}

\maketitle
%\linenumbers
\section{INTRODUCTION}
\label{sec: introduction}

The stochastic dynamics of defects plays central roles in a range of macroscopic phenomena---such as melting, sliding, and mobility---in condensed matter physics \cite{Reichhardt_2017, kim_dynamical_2020, Bechinger_2012, hasnain_sliding_2013, harada_vortex_1996, Cornell_BEC_2006}. In two-dimensional (2D) crystals, a cornerstone theory in this field deals with topological defects, the well-established Kosterlitz-Thouless-Halperin-Nelson-Young (KTHNY) theory \cite{Kosterlitz_1973, Kosterlitz_2016, Halperin_1978, Young_1979} which describes a two-stage melting transition, driven by the sequential unbinding of topological defect pairs. As is verified in colloidal and vortex lattice systems \cite{Zahn_1999, Roy_2019}, it establishes that the proliferation of first dislocations and then disclinations drives the system from a crystal to a hexatic fluid and finally to an isotropic liquid. Point defects in 2D crystals, such as vacancies and interstitials, can be viewed as clusters of dislocations (e.g., dislocation dipoles) \cite{kim_dynamical_2020}. Since dislocations themselves are topological defects, the motion of point defects are intimately connected to the broader category of topological defect behavior, suggesting their critical role in the initial stage of 2D melting. Indeed, point defects are fundamentally important for understanding the macroscopic properties of materials \cite{A0_1999, Guo_2022, Interstitials_1948, Interstitials_1949, Point_1984}.

Colloidal crystals have served as a model system for exploring a wealth of statistical phenomena and their underlying mechanisms \cite{phys_today_1998, Pertsinidis_2005}. They present an excellent platform for probing point-defect dynamics in 2D lattices. For instance, Ling \textit{et al.}~\cite{pertsinidis_diffusion_2001, kim_dynamical_2020} recorded the trajectories of point defects and the corresponding lattice distortions in such systems, advancing our understanding of their migration, including stable configurations and diffusion mechanisms. Subsequent computational studies \cite{Reichhardt_simulation_2007, Oliveira_2007, Oliveira_mechanism_2011} on similar systems have examined the influence of key variables such as interaction strength between colloids and the surrounding temperature. A recent experiment \cite{kim_dynamical_2020} has further revealed that the movement of a defect consisting of dual interstitials may violate detailed balance, leading to local lattice melting. These advances are achieved with experimental techniques~\cite{Pertsinidis_2005} such as creating artificial defects and tracking the evolution of lattice distortions and the trajectories of the point defects.

Nevertheless, the analyses of point-defect trajectories have so far been largely interpreted within the paradigm of a simple diffusive model---namely, ordinary Brownian motion characterized by zero drift and a constant diffusion coefficient---which portrays the motion of individual colloidal particles driven by thermal fluctuations in an isotropic fluid environment \cite{phys_today_1998}. In reality, the movement of a point defect is a collective, multi-body phenomenon under the background of an anisotropic crystalline lattice. Therefore, defect trajectories may exhibit dynamics far more complex than the basic model. An enhanced study of these paths may be essential, as they constitute the fundamental data for probing dynamical processes. The context raises naturally a pivotal question: Do experimental time-series trajectories of point defects harbor richer dynamical information than previously recognized? In other words, can we extract deeper knowledge of the defect dynamics from the set of primitive data?

Theoretically, a framework for stochastic dynamics termed ``evolution mechanics'' as proposed in \cite{Ao_2005}, has been applied across diverse fields including physics, biology, medicine and artificial intelligence \cite{kwon_structure_2005, Yin_2006, ao_existence_2007, Ao_2008, Kwon_2011, Yuan_2012, shi_relation_2012, Ao_dynamical_2013, Tang_2013, Mayian_2013, XiongXia_2023}. The formalism describes the time evolution of a complex system using equations of motion of generic variables, analogous to the Langevin equations in statistical physics for stochastic trajectories in the phase space. The formalism can be decomposed into two dynamical components under a general stochastic potential: a dissipative component, determined by the diffusion matrix of the underlying generalized Brownian motion, and a transverse component that conserves the ``energy'' of the potential. An equivalent Fokker-Planck equation can be derived, whose steady state (when it exists) follows a Boltzmann-Gibbs-like distribution shaped by the stochastic potential \cite{Yin_2006, Ao_2008}.

In this work, we demonstrate that the aforementioned motion data contain rich dynamical signatures amenable to quantitative extraction. Under the evolution mechanics framework, we infer the stochastic potential landscape directly from the trajectory data. It uncovers energy differences between local minima that align with experimental observations, and yields an activation energy for point-defect dynamics on the order of $10^{-3}$ to $10^{-2}$ times the crystal binding energy, consistent with prior estimates for this system \cite{Pertsinidis_2005}. Significant results are obtained from the analysis of individual trajectories corresponding to four distinct types of point defects.

The remainder of this paper is structured as follows. Sec.~\ref{sec: Methods} begins with detailing the source and interpretation of the experimental time-series data. We next outline the framework of evolution mechanics, focusing on the aspects central to this study. It is then applied to the point-defect dynamics, along with the core mathematical methods. Sec.~\ref{sec: Results} presents the results, including that of the drift and diffusion terms, the derived stochastic potential, and the simulations based on the constructed SDEs. We conclude with a discussion in Sec.~\ref{sec: Discussion}, relating our findings to the broader context of studying dynamics in complex systems and providing concluding remarks on future directions. Additionally, some excess mathematical algebra is placed in Appendix at the end of the paper.

\section{Methods}
\label{sec: Methods}

\subsection{Data Source and Description}

This work focuses on the experimental data from Ling \textit{et al.}~\cite{pertsinidis_diffusion_2001, kim_dynamical_2020}, which characterized four distinct types of point defects in a 2D hexagonal colloidal crystal: mono- and di-vacancies (i.e. dual vacancies) \cite{pertsinidis_diffusion_2001}, and mono- and di-interstitials (dual interstitials) \cite{kim_dynamical_2020}. The 2D colloidal crystals were formed via self-assembly of sulfate-polystyrene microspheres (diameter $0.3 \,\si{\micro\meter}$) in a highly deionized aqueous solution at room temperature ($T = 295 \,\mathrm{K}$). The system was confined between two parallel substrates separated by approximately $2 \,\si{\micro\meter}$, resulting in a single-layer hexagonal lattice with a lattice constant of about $1.1 \,\si{\micro\meter}$. The measured binding energy of the crystal was $346 \,k_BT$, confirming that the system is in a deep solid phase where thermally generated defects are negligible \cite{Pertsinidis_2005}. Point defects were artificially introduced into the pristine lattice using optical tweezers, after which they migrated independently. Afterwards, the temporal evolution of each defect was tracked by defining its position as the center of mass of its constituent disclinations \cite{pertsinidis_diffusion_2001, kim_dynamical_2020}. This operational definition, employed in the original studies, allows for robust tracking by leveraging the topological structure of the defects. Recording the instantaneous lattice configuration over time gives rise to the time-series trajectories of the point defects in question.

\begin{figure*}[t]
    \centering
    \subfigure[Mono-vacancy]{
        \includegraphics[width=0.22\textwidth]{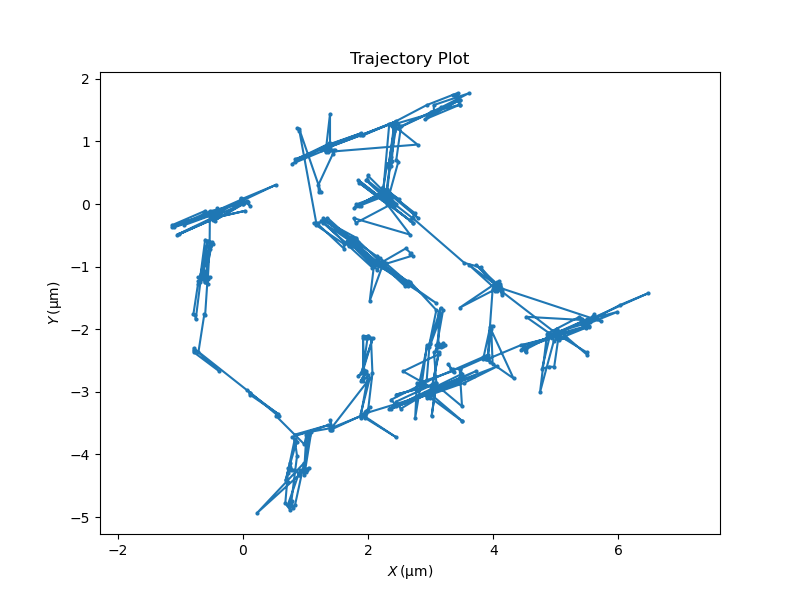}
    \label{1a}
    }
    \subfigure[Di-vacancy]{
        \includegraphics[width=0.22\textwidth]{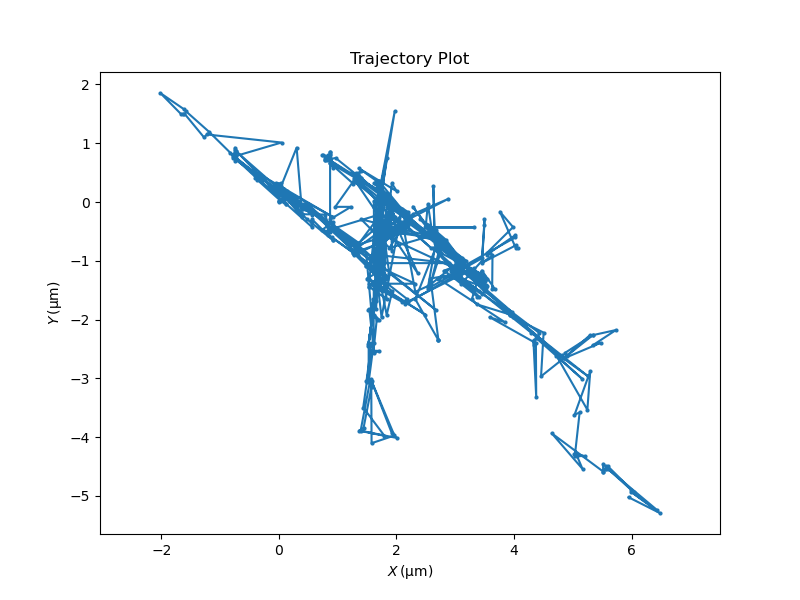}
    \label{1b}
    }
    \subfigure[Mono-interstitial]{
        \includegraphics[width=0.22\textwidth]{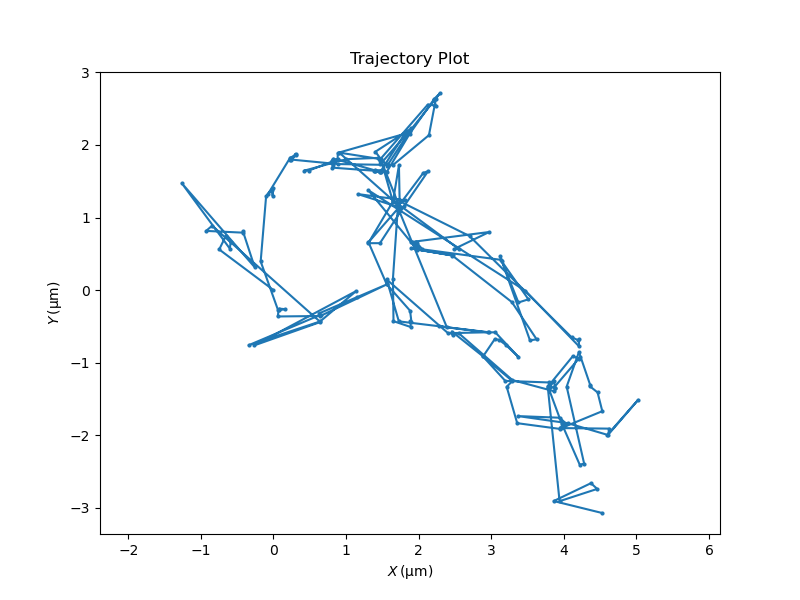}
    \label{1c}
    }
    \subfigure[Di-interstitial]{
        \includegraphics[width=0.22\textwidth]{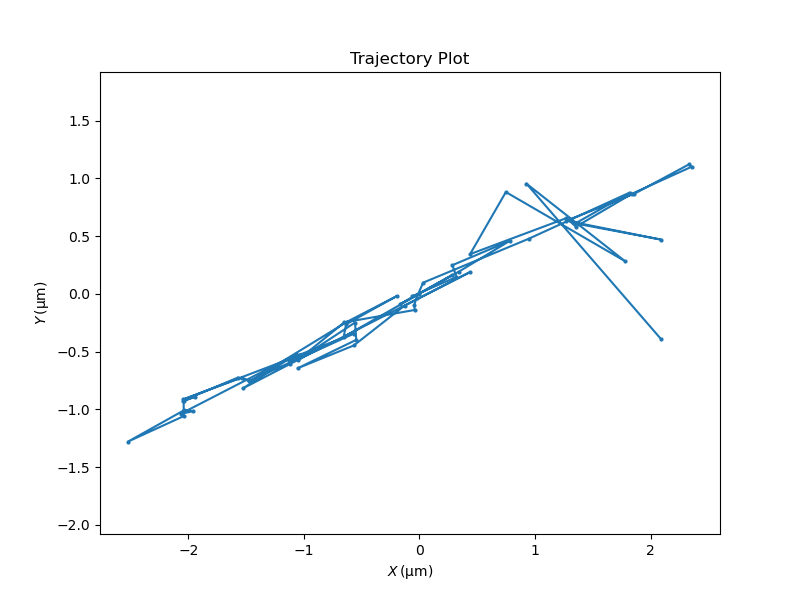}
    \label{1d}
    }
\caption{
Trajectories of vacancies and interstitials.
Panels (a)–(d) are extracted from the time series data obtained from videos; they represent a subset of the trajectories originally published as FIG.~2 in Ref.~\cite{pertsinidis_diffusion_2001} and FIG.~3 in Ref.~\cite{kim_dynamical_2020}.
The dots represent the center-of-mass positions of the disclinations.
}
\label{F1}
\end{figure*}

We quantified the dynamics using time-resolved positional data extracted from the publicly available video datasets associated with Refs.~\cite{pertsinidis_diffusion_2001, kim_dynamical_2020}. The specific datasets used in this study are as follows: mono-vacancy ($600$ frames at 60 fps), di-vacancy (600 frames at 60 fps), mono-interstitial (197 frames at 30 fps), and di-interstitial (61 frames at 30 fps).
The overall paths traversed by these point defects, shown in FIG.~\ref{F1}, exhibit a visual resemblance to the paths of Brownian particles.
The qualitative observation strongly supports the notion of point defects as quasi-particles, an abstraction that has proven fruitful in prior studies of vortex dynamics \cite{A0_1999, Guo_2022}. This quasi-particle picture provides the foundational motivation for analyzing their dynamics within a comprehensive stochastic framework.

\subsection{Overview of Evolution Mechanics}

We first present an overview of the so-called evolution mechanics, focusing on the concepts and mathematical relations essential for its application to point-defect dynamics. The discussion is restricted to the time-homogeneous case. For clarity, vectors and matrices are denoted in boldface throughout the remainder of this work.

\subsubsection{Fundamental Equations of Stochastic Dynamics}

Consider a general complex system characterized by a continuous $N$-dimensional state vector $\mathbf{q}$. Within the framework of evolution mechanics, the dynamics of such a system may be characterized by three mathematically equivalent formulations of its equations of motion \cite{Ao_2005, Yin_2006, Yuan_2012}:

1. \textbf{Stochastic Differential Equation (SDE)}
\begin{subequations}
\label{SDE_both}
\begin{align}
\label{SDE_form}
\dot{\mathbf{q}}(t) &= \,\mathbf{f}(\mathbf{q}) + \bm{\xi}(\mathbf{q},t), \\
\label{fluc-diss-relation-1}
\left\langle \bm{\xi}(\mathbf{q},t)\bm{\xi}^{T}(\mathbf{q},t') \right\rangle &= 2{\epsilon}\mathbf{D}(\mathbf{q})\,\delta(t-t').
\end{align}
\end{subequations}
where $\dot{\mathbf{q}}$ denotes the time derivative of ${\mathbf{q}}(t)$ and the superscript $T$ indicates the transpose. In Eq. (\ref{SDE_both}), the vector $\mathbf{f}(\mathbf{q})$ represents the deterministic drift term, and the stochastic component is modeled by a Gaussian white noise vector $\bm{\xi}(\mathbf{q},t)$ with zero mean and a covariance given by Eq.~\eqref{fluc-diss-relation-1}. Here $\delta(\tau)$ is the Dirac delta function, indicating the noise is uncorrelated in time; the matrix $\mathbf{D}(\mathbf{q})$ is a symmetric and semi-positive definite diffusion matrix; and the angle bracket pair $\langle...\rangle$ refers averaging over the noise distribution. Both the drift and diffusion terms exhibit no explicit time dependence, consistent with our restricted scope to the time-homogeneous case only. The positive constant $\epsilon$ sets the noise intensity; its physical interpretation will be clarified in connection with Eq.~\eqref{Bolzmann-Gibbs-like-distribution} below.

2. \textbf{A Canonical Equivalence}
\begin{subequations}
\begin{align}
\label{canonical_form}
[\mathbf{S}(\mathbf{q}) + \mathbf{A}(\mathbf{q})] \dot{\mathbf{q}} &= \, -\nabla\phi(\mathbf{q}) + \bm{\zeta}(\mathbf{q}, t),  \\
\label{fluc-diss-relation-2}
\left\langle \bm{\zeta}(\mathbf{q},t)\bm{\zeta}^{T}(\mathbf{q},t') \right\rangle &= 2{\epsilon}\mathbf{S}(\mathbf{q})\,\delta(t-t'),
\end{align}
\end{subequations}
where $\nabla$ is the gradient operator in the state space. The scalar function $\phi(\mathbf{q})$ is referred to as the stochastic potential in the current study. The symmetric, semi-positive definite friction matrix $\mathbf{S}(\mathbf{q})$ governs dissipative dynamics along the negative gradient of $\phi(\mathbf{q})$, while the antisymmetric matrix $\mathbf{A}(\mathbf{q})$ drives non-dissipative, transverse motion on iso-potential surfaces. In the zero-noise limit ($\epsilon\to0^+$), the two matrices ensure the value of the potential being monotonically decreasing over time, making the latter a Lyapunov function for the dynamics.  The zero-mean Gaussian white noise $\bm{\zeta}(\mathbf{q}, t)$ is related to the friction matrix by the fluctuation-dissipation relation given in Eq.~\eqref{fluc-diss-relation-2}. Eqs.~\eqref{fluc-diss-relation-1} and \eqref{fluc-diss-relation-2} both represent the fluctuation-dissipation theorem.

3. \textbf{Fokker-Planck Equation (FPE)}
\begin{equation}
\label{FPE_form}
\partial_{t} \rho(\mathbf{q}, t) = \nabla \cdot
\left\{
[\mathbf{D}(\mathbf{q})+\mathbf{Q}(\mathbf{q})]\left[ \epsilon\nabla + (\nabla\phi(\mathbf{q})) \right]
\right\}
\rho(\mathbf{q},t),
\end{equation}
where $\rho(\mathbf{q}, t)$ is the probability density function of the state $\mathbf{q}$ at time $t$, and $\partial_t$ denotes the partial derivative with respect to time, and the antisymmetric matrix $\mathbf{Q}(\mathbf{q})$ is given by Eq.~(\ref{diss-diffu-relation}) below. This form of the Fokker-Planck equation highlights the role of the stochastic potential $\phi(\mathbf{q})$. When an steady state (including equilibrium) is reached, it takes the simple form of a Boltzmann-Gibbs-like distribution:
\begin{equation}
\label{Bolzmann-Gibbs-like-distribution}
\rho(\mathbf{q}; t = +\infty) \propto \exp\left(- \frac{\phi(\mathbf{q})}{\epsilon} \right),
\end{equation}
which reveals that the parameter $\epsilon$ acts as an effective temperature to the dynamics. In other words, $\phi/\epsilon$ corresponds to the ``thermal activation energy" in units of $k_\mathrm{B}T$ for a system in equilibrium.

The FPE given in Eq.~\eqref{FPE_form} is derived within the evolution mechanics through methods such as the zero-mass limit~\cite{Yin_2006}, rather than from a pre-defined stochastic integral. Therefore, the mathematical equivalence between this FPE and the SDE presented in Eqs.~\eqref{SDE_form} and \eqref{FPE_form} underlines the existence of a corresponding stochastic integral that is distinct from the conventional It\^{o} or Stratonovich calculus \cite{ao_existence_2007}. For a detailed discussion of the relationship between these stochastic integrals and their corresponding FPEs, we refer the reader to Refs.~\cite{Yuan_2012, shi_relation_2012}.

The various matrices involved in the descriptions are related through the following fundamental equation
\begin{equation}
\label{diss-diffu-relation}
[\mathbf{S}(\mathbf{q}) + \mathbf{A}(\mathbf{q})]^{-1} = \mathbf{D}(\mathbf{q}) + \mathbf{Q}(\mathbf{q}).
\end{equation}
Clearly, a non-zero transverse matrix $\mathbf{Q}$ implies a non-zero $\mathbf{A}$ and vice versa. In the one-dimensional case, the antisymmetric matrices vanish ($\mathbf{Q}=\mathbf{A}=0$) so that Eq.~\eqref{diss-diffu-relation} reduces to the familiar Einstein relation, $SD=1$.

\subsubsection{Dynamical Structure Decomposition (DSD)}

Eq.~\eqref{SDE_form}, a standard SDE, decomposes the dynamics into a deterministic component, $\mathbf{f}(\mathbf{q})$, and a stochastic component, $\bm{\xi}(t)$ \cite{Ao_2005, Gardiner_handbook_2004}. These components can, in principle, be separated from experimental trajectories, as outlined in Eq.~\eqref{two_terms} below. This formulation thus provides a crucial link between empirical trajectory data and theoretical description. The stochastic component $\bm{\xi}(t)$ is a Gaussian white noise with zero mean, and its statistical properties are completely characterized by the diffusion matrix $\mathbf{D}(\mathbf{q})$ via the fluctuation-dissipation relation in Eq.~\eqref{fluc-diss-relation-1}. Consequently, knowledge of both the drift vector $\mathbf{f}(\mathbf{q})$ and the diffusion matrix $\mathbf{D}(\mathbf{q})$ in Eq.~\eqref{SDE_both} provides a comprehensive description of the stochastic dynamics. The initial step in applying evolution mechanics is naturally the computation of these two quantities from the data.

When the time difference $t'-t$ is an infinitesimal $\mathrm{d}t$, Eq.~\eqref{fluc-diss-relation-1} implies the relation $\left\langle \bm{\xi}(\mathbf{q},t)\bm{\xi}^{T}(\mathbf{q},t + \mathrm{d}t) \right\rangle=\mathbf{D}(\mathbf{q})/\mathrm{d}t$ (hereafter, for simplicity and without loss of generality, we set the noise intensity $\epsilon = 1/2$) \cite{Gardiner_handbook_2004}. Using this relation, the drift vector and diffusion matrix can be expressed in terms of the first and second conditional moments of the state variable $\mathbf{q}$:
\begin{subequations}
\label{two_terms}
\begin{align}
\mathbf{f}(\mathbf{q}_0) &= \lim_{\tau \to 0^+} \frac{1}{\tau} \langle \Delta\mathbf{q} \rangle|_{\mathbf{q}(0)= \mathbf{q}_0} \,, \\
\mathbf{D}(\mathbf{q}_0) &= \lim_{\tau \to 0^+} \frac{1}{\tau} \left.\left\langle (\Delta\mathbf{q}) (\Delta\mathbf{q})^{T} \right\rangle\right|_{\mathbf{q}(0) = \mathbf{q}_0} \,.
\end{align}
\end{subequations}
In these expressions, $\Delta\mathbf{q} \equiv \mathbf{q}(\tau) - \mathbf{q}(0)$, $\tau$ is a time interval, $\mathbf{q}_0$ is the state vector at time $t_0=0$, and $\langle...\rangle$ denotes averaging over the noise distribution. Eq.~\eqref{two_terms} provides a direct method for computing the drift and diffusion terms from a trajectory $\{\mathbf{q}(t)\}$, a method that has proven effective in practice \cite{FRIEDRICH_2000, nabeel_discovering_2025}.

The canonical form in Eq.~\eqref{canonical_form} decomposes the stochastic motion into three fundamental components: dissipative dynamics, conservative (non-dissipative) dynamics, and the stochastic potential $\phi$.
The equivalence between Eqs.~\eqref{SDE_form} and \eqref{canonical_form} suggests the existence of a stochastic potential $\phi$ for any process described by an SDE. After extracting $\mathbf{f}$ and $\mathbf{D}$ from trajectories, a central task remains: How can we decompose the dynamics according to Eq.~\eqref{canonical_form}, specifically, can we obtain the stochastic potential $\phi(\mathbf{q})$?

Combining Eq.~\eqref{canonical_form} with the matrix relation in Eq.~\eqref{diss-diffu-relation} and comparing with Eq.~\eqref{SDE_form} yields an expression for the drift vector \cite{Ao_2005, Yuan_2012}, the so-called dynamical structure decomposition (DSD)
\begin{equation}
\label{dynamical_structure_decomposition}
\mathbf{f}(\mathbf{q}) = - [\mathbf{D}(\mathbf{q}) + \mathbf{Q}(\mathbf{q})] \nabla \phi(\mathbf{q}).
\end{equation}
Here the drift becomes the superposition of two distinct dynamical modes governed by the symmetric diffusion matrix $\mathbf{D}$ and the antisymmetric matrix $\mathbf{Q}$.
From a technical standpoint, however, the DSD equation alone is insufficient to uniquely determine $\phi(\mathbf{q})$ because the terms directly extracted from the SDE are $\mathbf{f}$ and $\mathbf{D}$ [given by Eq.~(\ref{fluc-diss-relation-1})], with $\mathbf{Q}$ remaining unknown.
Consequently, for a system with $\mathbf{Q}$-dynamics (i.e., where $\mathbf{Q}$ is non-zero), constructing the stochastic potential is less straightforward. For systems that can be linearized around a fixed point, a rigorous general method to determine $\phi(\mathbf{q})$ via the treatment of $\mathbf{Q}$ was established in 2005~\cite{kwon_structure_2005}. In nonlinear regimes, however, no general analytical method exists to determine the potential $\phi(\mathbf{q})$. In such cases, one must typically resort to numerical techniques or approximation schemes---such as the gradient expansion method \cite{Ao_2004}---to obtain an approximate potential.

To summarize, the primary motivation of evolution mechanics is to transform a complex system into a generic stochastic model that conventional concepts of statistical physics can be adopted. To this end, the stochastic potential $\phi(\mathbf{q})$ plays a dual role. First, as seen from the DSD in Eq.~\eqref{dynamical_structure_decomposition}, its gradient $-\nabla{\phi}(\mathbf{q})$ directly determines the deterministic part of the drift, thereby governing the average dynamics. Second, in accordance with the Boltzmann-Gibbs-like distribution in Eq.~\eqref{Bolzmann-Gibbs-like-distribution}, the local minima of $\phi(\mathbf{q})$ identify the most probable positions in the steady state.

\subsection{Application to Point Defects}

The dynamics of a point defect, when modeled as a quasi-particle propagating on a hexagonal lattice, can be described by  stochastic motion of its position vector $\mathbf{q}(t)$ and governed by the time-homogeneous SDE in Eq.~\eqref{SDE_both}. This subsection details the specific analytical methods applied within the evolution mechanics framework. These methods form the basis for the results presented in Sec.~\ref{sec: Results}.

\subsubsection{Construction of Drift and Diffusion via Time Series}

Eq.~\eqref{two_terms} provides a method to derive the drift vector and diffusion matrix directly from discrete time-series data $\{\mathbf{q}_i\}$. For a chosen analysis time interval $\tau > 0$, which is an integer multiple of the data sampling interval, the continuum expressions are approximated by:
\begin{subequations}
    \label{discrete_approximations_of_two_terms}
\begin{align}
\mathbf{f}(\mathbf{q}_0) &\approx \frac{1}{\tau} \langle \Delta\mathbf{q} \rangle|_{\mathbf{q}(0)= \mathbf{q}_0} \,, \\
\mathbf{D}(\mathbf{q}_0) &\approx \frac{1}{\tau} \left.\left\langle [\Delta\mathbf{q} - \tau\,\mathbf{f}(\mathbf{q}_0)] [\Delta\mathbf{q} - \tau\,\mathbf{f}(\mathbf{q}_0)]^{T} \right\rangle\right|_{\mathbf{q}(0) = \mathbf{q}_0} \,.
\end{align}
\end{subequations}
To compute these averages, we evaluate them at nodes of a spatial grid. For a node at position $\mathbf{q}_0$, the averages include all trajectory segments where the initial point falls within a spatial bin of radius $r$ centered at $\mathbf{q}_0$. This spatial averaging is essential for accumulating sufficient statistics. %Only estimates of $\mathbf{D}(\mathbf{q}_0)$ that yield a positive semi-definite matrix are retained in the subsequent analysis.
Clearly, $\mathbf{D}(\mathbf{q}_0)$ so estimated yields a semi-positive definite matrix as it should.

The accuracy of these approximations are critically sensitive to the choice of the spatial bin size $r$ and the analysis time interval $\tau$. The spatial scale $r$ represents a trade-off: it must be large enough to ensure statistical reliability ($r> r_{\mathrm{min}}$) yet small enough to resolve spatial variations in the dynamics ($r< r_{\mathrm{max}}$). The choice of $\tau$ is constrained by two factors. It must be large enough that the average displacement is significant relative to the measurement noise ($\tau>\tau_{\mathrm{min}}$), and small enough to characterize the locality of $\mathbf{f}(\mathbf{q}_0)$ and $\mathbf{D}(\mathbf{q}_0)$. Consequently, reliable results are obtained within an intermediate range for both parameters. The specific boundaries of this range are determined by the properties of each trajectory dataset. The native sampling rate (e.g., $60$ fps for vacancies, $30$ fps for interstitials) sets a lower bound on the practically resolvable $\tau$, while the relevant dynamical timescales of the system determine the optimal value.

\subsubsection{Statistical Characterization: Averages and Position Dependence}

To enable comparison with prior studies that reported scalar diffusion constants under the assumption of ordinary Brownian motion, we first compute weighted averages of the constructed dynamical terms. Specifically, we evaluate the average magnitude of the drift vector and the average eigenvalues of the diffusion matrix, which are summarized in Table \ref{table:diffusion_comparison}. All averaging, and indeed all optimizations in this work, employ a weighting scheme where the contribution from each spatial grid point is proportional to the number of individual positional measurements within its associated spatial bin. This approach ensures that regions with higher sampling density contribute more significantly to the aggregate statistics.

To quantitatively assess whether the drift and diffusion terms exhibit positional dependence beyond a constant model, we employ the coefficient of determination, denoted as $R^2$. This metric compares the goodness-of-fit between a simple constant model (no positional variation) and a specified parametric model. The $R^2$ statistic is defined as:
\begin{equation}
\label{R2}
R^2 = 1 - \frac{\sum_{i=1}^{n} w_i (y_i - \hat{y}_i)^2}{\sum_{i=1}^{n} w_i (y_i - \bar{y}_w)^2},
\end{equation}
where $n$ is the number of spatial grid points with valid estimates of the dynamical term $y_i$ (e.g., a specific component of the drift vector or an element of the diffusion matrix). Here, $\hat{y}_i$ is the value predicted by the specified model, $\bar{y}_w = \left({\sum_{i=1}^n w_iy_i}\right)/\left({\sum_{i=1}^n w_i}\right)$ is the weighted mean, and $w_i$ is the weight assigned to the $i$-th grid point (proportional to its sample count).
Here, the specified model for calculating \(R^2\) is a periodic model similar to Eq.~\eqref{Fourier_series_expansion}, whose period matches that of the 2D hexagonal lattice and whose parameters are estimated via least-squares fitting. It should be noted that this model is used only for the \(R^2\) computation and is not presumed in other calculations unless otherwise stated.

The $R^2$ value quantifies the proportion of the weighted variance in the dynamical term that is explained by the specified model relative to the constant model. An $R^2$ value close to zero indicates that the constant model is sufficient, suggesting no strong evidence for positional dependence. Conversely, a higher $R^2$ value reflects a stronger agreement with the specified periodic description, indicating that the dynamical term varies systematically with position within the lattice.

\subsubsection{Periodic Stochastic Potential via Least-Squares Fitting}

Given the limited volume of trajectory data available in the present study, we adopt a diffusive approximation by setting $\mathbf{Q}=0$ in the DSD equation [Eq.~\eqref{dynamical_structure_decomposition}] for the construction of the stochastic potential. This simplification implies that we consider only the symmetric, dissipative component of the dynamics (captured by the diffusion tensor $\mathbf{D}(\mathbf{q})$) while neglecting the possible antisymmetric, transverse forces represented by $\mathbf{Q}(\mathbf{q})$. This approximation is justified on two primary grounds. First, it is exact in the scenario where the underlying system dynamics are inherently non-$\mathbf{Q}$ in nature. Second, and more pragmatically, it circumvents significant technical difficulties inherent in the general case. A faithful decomposition of the drift field $\mathbf{f}(\mathbf{q})$ requires the simultaneous determination of both the potential $\phi(\mathbf{q})$ and the spatially varying matrix $\mathbf{Q}(\mathbf{q})$. In principle, with sufficient data, one could determine $\mathbf{Q}(\mathbf{q})$---first approximated as a constant matrix and then iteratively refined to account for its positional dependence. However, the determination of $\mathbf{Q}(\mathbf{q})$ constitutes a nonlinear inverse problem that is ill-posed with limited data. But the dataset in this study does not warrant such an advanced and potentially unstable analysis. As a result, we proceed with the diffusion limit ($\mathbf{Q}\cong 0$), acknowledging that a comprehensive investigation of a finite $\mathbf{Q}(\mathbf{q})$ remains an important topic for future work.

The stochastic potential $\phi(\mathbf{q})$, whose gradient determines the deterministic part of the dynamics (i.e., the drift term) through the DSD relation in Eq.~\eqref{dynamical_structure_decomposition}, is defined on the state space of the defect position vector $\mathbf{q}$. Given that the point defects move within a crystal possessing a 2D hexagonal lattice,
and as a first-order approximation we focus on the dominant periodicity while neglecting secondary effects such as hydrodynamic interactions,
we postulate that the stochastic potential $\phi(\mathbf{q})$ itself inherits the periodicity of the underlying lattice. This postulate provides the necessary constraint to construct the continuous potential function from the discrete values of the drift and diffusion terms, which were in turn extracted from the experimental trajectories. Note that this construction imposes periodicity only on the potential $\phi$, and it has not yet presumed any specific functional form for the drift and diffusion terms themselves.

Within the reciprocal lattice appropriate for the 2D hexagonal crystal structure \cite{Kittel_introduction_2004}, the periodic stochastic potential is expressed as a Fourier series:
\begin{equation}
\label{Fourier_series_expansion}
\phi(\mathbf{q}) = \sum_{\mathbf{G}} V_{\mathbf{G}} \exp(i\mathbf{G}\cdot\mathbf{q}),
\end{equation}
where the summation runs over reciprocal lattice vectors $\mathbf{G}=m_1\mathbf{b}_1 + m_2\mathbf{b}_2$, with $\mathbf{b}_1$ and $\mathbf{b}_2$ being the primitive vectors of the reciprocal lattice corresponding to the hexagonal symmetry. For real  $\phi(\mathbf{q})$, we have $V^*_{\mathbf{G}}=V_{-\mathbf{G}}$ and set $V_0=0$ for convenience. In practice, the summation in Eq.~\eqref{Fourier_series_expansion} is truncated to include reciprocal lattice vectors up to a specific cutoff shell (a parameter defining the maximum distance $|\mathbf{G}|_{\mathrm{max}}$ in reciprocal space).

For $\phi(\mathbf{q})$ from Eq.~\eqref{Fourier_series_expansion}, we can define a residual at each measured position $\mathbf{q}_k$. In the diffusive limit, i.e. $\mathbf{Q}\cong 0$, the DSD relation of Eq.~\eqref{dynamical_structure_decomposition} simplifies to $\mathbf{f}(\mathbf{q}) = -\mathbf{D}(\mathbf{q})\nabla\phi(\mathbf{q})$. The residual vector $\mathbf{z}(\mathbf{q}_k)$ quantifies the deviation from this relation and is defined as:
\begin{equation}
\label{residual}
\mathbf{z}\left(\mathbf{q}_k; \{V_{\mathbf{G}}\}\right)
=
\mathbf{f}(\mathbf{q}_k) + \mathbf{D}(\mathbf{q}_k)\nabla\phi(\mathbf{q}_k; \{V_{\mathbf{G}}\}).
\end{equation}
As a result, the optimal Fourier coefficients $\{V_{\mathbf{G}}\}$ can be obtained by minimizing the overall of the residuals across all measured positions,
\begin{equation}
\label{SSR}
W = \sum_k w_k \, \mathbf{z}^T(\mathbf{q}_k)\mathbf{z}(\mathbf{q}_k),
\end{equation}
where the summation includes the previously defined weights $w_k$ proportional to the sampling density at $\mathbf{q}_k$.

Substituting the Fourier series expansion into the definition of $W$ yields an expression that is quadratic in the complex coefficients $\{V_{\mathbf{G}}\}$. Since constraint $V^*_{\mathbf{G}}=V_{-\mathbf{G}}$, only half of them can be regarded as independent, a situation analogous to handling complex fields in wave-vector space in electrodynamics \cite{schwabl2008advanced}. Hence certain care must be taken with respect to the standard least-squares procedure of setting partial derivatives $\partial W/\partial{V_{\mathbf{G}}}$ to zero. As detailed in Appendix, this leads to a system of linear equations for the coefficients $\{V_{\mathbf{G}}\}$,
\begin{equation}
\label{matrix_equation}
\sum_{\mathbf{G}'}A_{\mathbf{G}\mathbf{G}'}
V_{\mathbf{G}'} + C_{\mathbf{G}} = 0
\end{equation}
for all $\mathbf{G}$ within the cutoff. The coefficients $A_{\mathbf{G}\mathbf{G}'}$ and $C_{\mathbf{G}}$ are given by:
\begin{subequations}
\label{element_of_A_C}
\begin{align}
A_{\mathbf{G}\mathbf{G}'} &= \sum_{k} w_k \left(\mathbf{G}^{T}\mathbf{D}^{T}\mathbf{D}\mathbf{G}'\right) \exp\left[-i \,(\mathbf{G}-\mathbf{G}') \cdot \mathbf{q}_k\right] , \\
C_{\mathbf{G}} &= \sum_{k} w_k (-i\,\mathbf{f}^T\mathbf{D}\mathbf{G}) \exp(-i\,\mathbf{G}\cdot \mathbf{q}_k) .
\end{align}
\end{subequations}
where the summations on the $k$'s is over the spatial grid points and the solution yields the set of coefficients that minimize $W$.

\subsubsection{Optimal Points on the Stochastic Potential}
\label{subsubsec: potential}

The local minima and saddle points of this stochastic potential $\phi(\mathbf{q})$ can provide crucial dynamical or thermodynamic information. The former locate the locally steady points while the latter identify the transition paths between them. To locate and characterize these critical points, we utilize the gradient and Hessian matrix of $\phi(\mathbf{q})$, derived from its Fourier series representation in Eq.~\eqref{Fourier_series_expansion},
\begin{equation}
\nabla\phi(\mathbf{q}) = \sum_{\mathbf{G}}
(i\mathbf{G}) V_{\mathbf{G}} \exp(i\mathbf{G} \cdot \mathbf{q}),
\end{equation}
\begin{equation}
\mathbf{H}(\mathbf{q}) = \nabla(\nabla\phi(\mathbf{q}))^T = -\sum_{\mathbf{G}} (\mathbf{G}\mathbf{G}^T)V_{\mathbf{G}}\exp(i\mathbf{G}\cdot\mathbf{q}).
\end{equation}
A local minima has a positive definite Hessian matrix, whereas a saddle point bas both mixed signs of eigenvalues for the matrix.

\subsubsection{SDE Construction and Numerical Integration}
\label{sec: Sim}

To simulate trajectories by integrating the SDE requires construction of the drift and diffusion terms, extending beyond the previously obtained stochastic potential $\phi(\mathbf{q})$. Although the prior method analyzes $\phi(\mathbf{q})$ without constraints on these terms, we impose a 2D hexagonal periodic condition on the diffusion matrix $\mathbf{D}(\mathbf{q})$ to enable a least-squares fit using an expansion similar to Eq.~\eqref{Fourier_series_expansion}. The fitted $\mathbf{D}(\mathbf{q})$ yields the drift term $\mathbf{f}(\mathbf{q})$ via Eq.~\eqref{dynamical_structure_decomposition} with $\mathbf{Q}\to0$. The noise term $\bm{\xi}(\mathbf{q}, t)$ in Eq.~\eqref{SDE_both} is then constructed from a lower triangular matrix $\mathbf{L}(\mathbf{q})$, which is obtained from the Cholesky decomposition that satisfies $\mathbf{D}(\mathbf{q}) = \mathbf{L}(\mathbf{q})\mathbf{L}^T(\mathbf{q})$. This additional periodic constraint applies only after determining $\phi(\mathbf{q})$, leaving its analysis unaffected.

A practical challenge stems from the numerical stability of the Cholesky decomposition, which requires the matrix $\mathbf{D}(\mathbf{q})$ to be strictly positive definite at all positions. In data-sparse regimes---exemplified by our mono-vacancy simulations, where over 50\% of pre-fit matrices were only positive semi-definite because certain positions had merely one sample point---pointwise estimates frequently violate this requirement. The periodic-constrained fit introduced earlier simultaneously addresses this issue by exploiting the inherent smoothness of the field, raising the proportion of positive-definite matrices to approximately 96\%. Any remaining non-positive-definite matrices are projected onto the nearest positive-definite counterpart via eigenvalue clipping, i.e., $\lambda_i \to \max\{\lambda_i, \delta\}$ with $\delta = 10^{-8}$, thereby ensuring strict positivity throughout.

For numerical integration of the SDE, we use the Euler-Maruyama scheme consistent with the It\^{o} interpretation in Eq.~\eqref{discrete_approximations_of_two_terms}. The discretized form is $\mathbf{q}(t+\Delta t) \approx \mathbf{q}(t) + \mathbf{f}(\mathbf{q}(t)) \Delta t + \mathbf{L}(\mathbf{q}(t)) \Delta \mathbf{W}(t)$, where $\Delta \mathbf{W}(t)$ denotes Wiener increments.

\section{Results}
\label{sec: Results}

\subsection{Non-Ordinary Brownian Dynamics: Spatially Varying Drift and Diffusion}

The drift vector, $\mathbf{f}(\mathbf{q})$, and diffusion matrix, $\mathbf{D}(\mathbf{q})$,
were derived from the experimental time‑series data using Eq.~\eqref{discrete_approximations_of_two_terms}, a procedure that depends on the choice of the time interval $\tau$ and the spatial bin size $r$.
This process yields discrete estimates of these terms across a grid of spatial positions. The key statistical properties of these terms---assessed through the coefficient of determination $R^2$ defined in Eq.~\eqref{R2} and spatial averaging---provide clear evidence of their positional dependence and serve as the foundation for constructing the stochastic potential in the following subsection.

\begin{table}[t]
\centering
\caption{The range of $R^2$ values (in $\%$) obtained from fitting the drift and diffusion terms across the parameter ranges $\tau=0.03\text{--}0.3 ~\mathrm{s}$ and $r=0.1\,\si{\micro\meter}$.
The cutoff shell is chosen as six ($42$ points). For the di-interstitial system ($61$ frames), this model with $42$ points may risk overfitting.
}
\label{tab:R2_r1}
\sisetup{table-format=-2.1, table-number-alignment=center}
\begin{tabular}{@{} l cc ccc @{}}
\toprule
\multirow{2}{*}{Defect Type} &
\multicolumn{2}{c}{Drift term [$\mathbf{f}(\mathbf{q_j})$]} &
\multicolumn{3}{c}{Diffusion term [$\mathbf{D}(\mathbf{q_j})$]} \\
\cmidrule(lr){2-3} \cmidrule(l){4-6}
& \multicolumn{1}{c}{$f_x$} & \multicolumn{1}{c}{$f_y$} & \multicolumn{1}{c}{$D_{xx}$} & \multicolumn{1}{c}{$D_{yy}$} & \multicolumn{1}{c}{$D_{xy}$}  \\
\midrule
Mono-vacancy & $39\text{--}65$ & $49\text{--}65$ & $23\text{--}72$ & $43\text{--}84$ & $41\text{--}67$  \\
Di-vacancy & $69\text{--}80$ & $51\text{--}73$ & $67\text{--}85$ & $47\text{--}83$ & $64\text{--}83$  \\
Mono-interstitial & $46\text{--}66$ & $48\text{--}75$ & $26\text{--}58$ & $37\text{--}49$ & $22\text{--}66$ \\
Di-interstitial & N/A & N/A & N/A & N/A & N/A \\
\bottomrule
\end{tabular}
\end{table}

\begin{table}[t]
\centering
\caption{Same as Table \ref{tab:R2_r1}, except that the spatial bin parameter range $r=0.2\text{--}0.6\,\si{\micro\meter}$.}
\label{tab:R2_r2-6}
\sisetup{table-format=-2.1, table-number-alignment=center}
\begin{tabular}{@{} l cc ccc @{}}
\toprule
\multirow{2}{*}{Defect Type} &
\multicolumn{2}{c}{Drift term [$\mathbf{f}(\mathbf{q_j})$]} &
\multicolumn{3}{c}{Diffusion term [$\mathbf{D}(\mathbf{q_j})$]} \\
\cmidrule(lr){2-3} \cmidrule(l){4-6}
& \multicolumn{1}{c}{$f_x$} & \multicolumn{1}{c}{$f_y$} & \multicolumn{1}{c}{$D_{xx}$} & \multicolumn{1}{c}{$D_{yy}$} & \multicolumn{1}{c}{$D_{xy}$}  \\
\midrule
Mono-vacancy & $1\text{--}22$ & $1\text{--}21$ & $1\text{--}28$ & $1\text{--}24$ & $1\text{--}29$  \\
Di-vacancy & $2\text{--}37$ & $1\text{--}20$ & $2\text{--}39$ & $2\text{--}32$ & $2\text{--}31$  \\
Mono-interstitial & $1\text{--}19$ & $1\text{--}22$ & $2\text{--}29$ & $2\text{--}34$ & $1\text{--}26$ \\
Di-interstitial & N/A & N/A & N/A & N/A & N/A \\
\bottomrule
\end{tabular}
\end{table}

The spatial variation of the dynamical terms is quantitatively demonstrated by the $R^2$ values summarized in Tables \ref{tab:R2_r1} and \ref{tab:R2_r2-6}. For the mono-vacancy, di-vacancy, and mono-interstitial, the $R^2$ values significantly exceed zero across a range of parameters $(\tau, r)$, indicating that both the drift and diffusion terms deviate substantially from a constant model. This signifies that the defect migration is not only influenced by a statistically significant, location-dependent drift but is also driven by multiplicative noise \cite{Van_Kampen_abbreviated_2007}. Furthermore, the $R^2$ value systematically decreases as the spatial bin size $r$ increases---for instance, dropping from over $50\%$ at $r=0.1 \,\si{\micro\meter}$ to about $1\%$ at $r=0.6 \,\si{\micro\meter}$ for a typical $\tau$---consistent with the expected averaging out of finer spatial variations over larger scales. Although the limited data for the di-interstitial precludes a reliable statistical assessment, the consistent behavior observed in the other three defect types strongly suggests that its dynamics also exhibit spatial inhomogeneity.

\begin{table*}[t]
\centering
\caption{Comparison of two terms from different analytical approaches. \textbf{Previous studies} (top section) impose the ordinary Brownian motion model where the \textit{Drift term} is constrained to zero, and the \textit{Diffusion term} is a single, isotropic diffusion constant. \textbf{This work} (bottom section) infers the two terms from the trajectories, reporting the magnitude of the drift vector (spatially averaged) as the \textit{Drift term} and the two eigenvalues of the (spatially averaged) diffusion matrix as the \textit{Diffusion term}.
For vacancies, the parameters are: $r=0.10~\si{\micro\meter}, \tau=0.015~\text{s}$, and cutoff shell $=7$. For interstitials, the parameters are: $r=0.12~\si{\micro\meter}, \tau=0.03~\text{s}$, and cutoff shell $=9$.
The effective number of data points $N$ is in parentheses.}
\label{table:diffusion_comparison}
\setlength{\tabcolsep}{0.7em}
\renewcommand{\arraystretch}{1.2}
\begin{tabular}{@{} l c c @{}}
\toprule
\textbf{System and Analysis} & \textbf{Drift Term} & \textbf{Diffusion Term} \\
& \textbf{(\si{\micro\meter\per\second})} & \textbf{(\si{\micro\meter\squared\per\second})} \\
\midrule
\textbf{Previous Studies} & & \\
\quad Mono-vacancy (Ref.~\cite{pertsinidis_diffusion_2001}, $N\sim7200$) & 0 (fixed) & 3.92 \\
\quad Di-vacancy (Ref.~\cite{pertsinidis_diffusion_2001}, $N\sim7200$) & 0 (fixed) & 4.45 \\
\quad Mono-interstitial (Ref.~\cite{kim_dynamical_2020}, $N=180$) & 0 (fixed) & 10.95 \\
\quad Di-interstitial (Ref.~\cite{kim_dynamical_2020}, $N=180$) & 0 (fixed) & 7.53 \\
\cmidrule(r){1-3}
\textbf{Data construction (This Work)} & & \\
\quad Mono-vacancy ($N=600$) & 0.55 & 3.85, 3.11 \\
\quad Di-vacancy ($N=600$) & 0.89 & 3.37, 8.66 \\
\quad Mono-interstitial ($N=197$) & 0.80 & 1.89, 2.94 \\
\quad Di-interstitial ($N=61$) & 1.06 & 6.05, 0.17 \\
\bottomrule
\end{tabular}
\end{table*}

To contextualize our results within the existing literature, we compare our spatially averaged dynamical terms with the effective diffusion constants reported in the prior investigations of the experimental data \cite{pertsinidis_diffusion_2001, kim_dynamical_2020} from which our trajectories are sourced. Those earlier works, which interpreted the trajectories under the assumption of ordinary Brownian motion, employed the formula $D = \langle|\nabla\mathbf{q}|^2\rangle/\tau$. Our results, summarized in Table \ref{table:diffusion_comparison}, are the weighted spatial averages of the drift vector magnitude and the eigenvalues of the diffusion matrix. The finite magnitude of the averaged drift vector ($0.5\text{--}1 \,\si{\micro\meter\per\second}$) observed here, in a system with no external drift, primarily reflects the statistical uncertainty at our chosen parameters. The disparity between the two averaged eigenvalues of the diffusion matrix provides clear evidence of anisotropic diffusion. This anisotropy is particularly pronounced for di-vacancies and di-interstitials (difference $> 5 \,\si{\micro\meter\squared\per\second}$), while a moderate anisotropy (difference $\approx 1 \,\si{\micro\meter\squared\per\second}$) is observed for the mono-defects.

The comparison of diffusion coefficients reveals nuanced consistency. For vacancies, for which our investigation utilizes a substantial subset (600 points from a 10-second path at 60 fps) and the prior study \cite{pertsinidis_diffusion_2001} used an even larger dataset ($\sim7200$ points), the previously reported diffusion constant is close to our two averaged eigenvalues, showing broad concordance. This suggests that when sufficient data is available, the spatial averaging in our method yields a result commensurate with the temporal averaging of the simpler model. For the mono-interstitial, however, a significant discrepancy exists. Both our examination (197 points) and the prior analysis \cite{kim_dynamical_2020} (180 points from a 6-second track at 30 fps) relied on relatively small datasets. This, combined with the inherent spatial inhomogeneity of the dynamics, means the two studies likely probe different parts of the trajectory, leading to different average values. In contrast, the results for the di-interstitial are in reasonable alignment, suggesting that its diffusion anisotropy is relatively consistent along the trajectory, as qualitatively observed in FIG.~3(B) of Ref.~\cite{kim_dynamical_2020}.

In summary, our analysis reveals that point defect dynamics are not fully captured by the paradigm of ordinary Brownian motion, which assumes isotropic diffusion, a vanishing drift term, and additive noise. The observed spatially varying drift and diffusion terms indicate a more complex stochastic process. Therefore, while the conventional practice of calculating a scalar diffusion constant can provide a useful average measure in some cases, it may not fully capture the complexity of the dynamics. A more general framework, such as the one applied here, is necessary to elucidate the underlying physical principles.

\subsection{Energetics and Dynamics under Stochastic Potential}
\label{subsec: Energy}

\begin{figure*}[t]
    \centering
    \subfigure[Mono-vacancy (N=600)]{
        \includegraphics[width=0.23\textwidth]{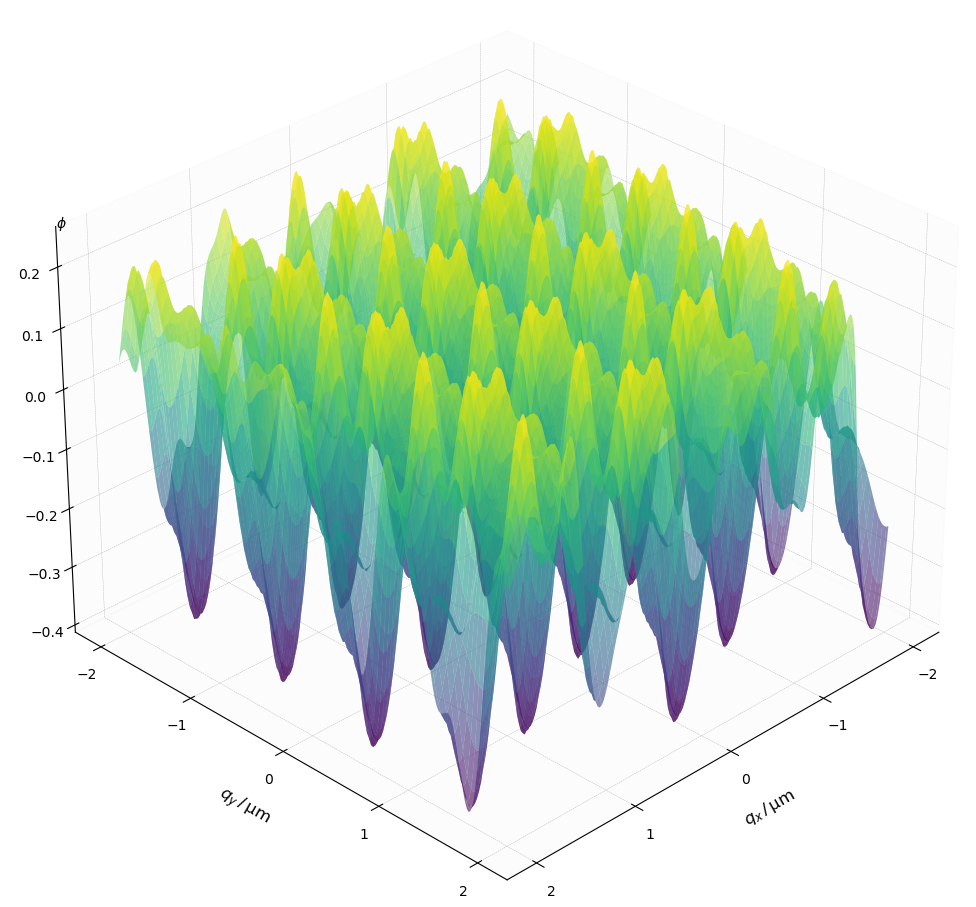}
        \label{3a}
    }
    \subfigure[Di-vacancy (N=600)]{
        \includegraphics[width=0.23\textwidth]{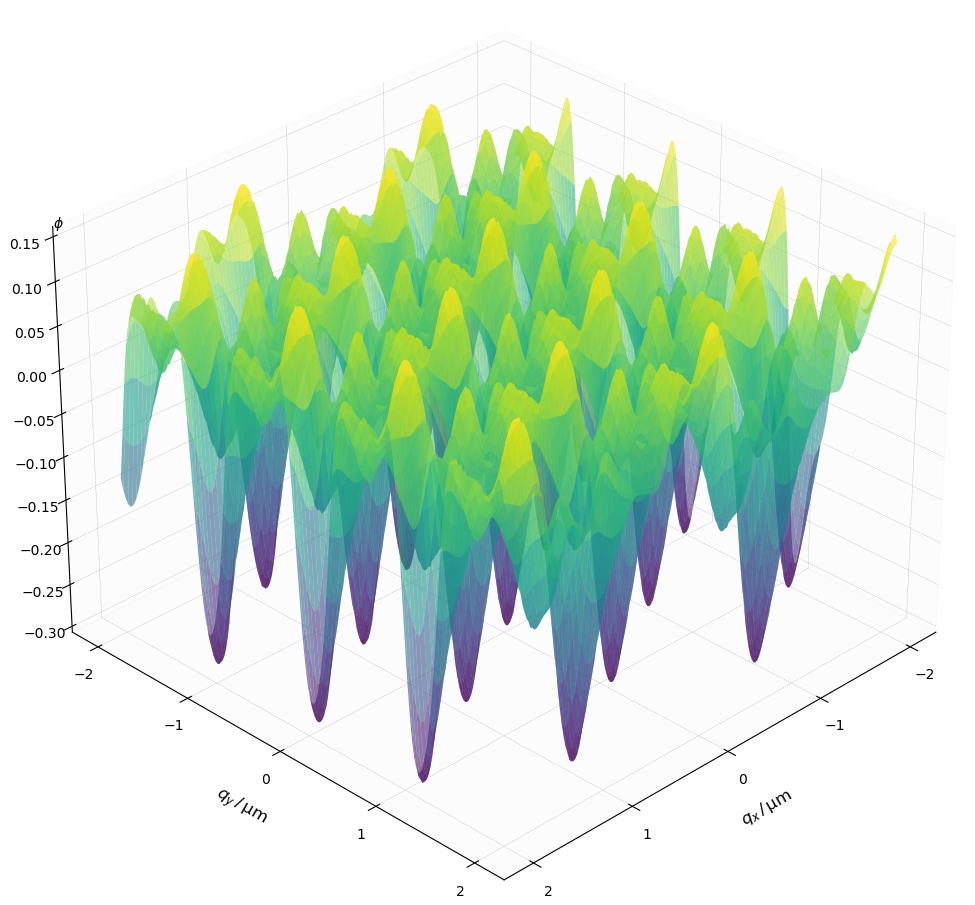}
    \label{3b}
    }
    \subfigure[Mono-interstitial (N=197)]{
        \includegraphics[width=0.23\textwidth]{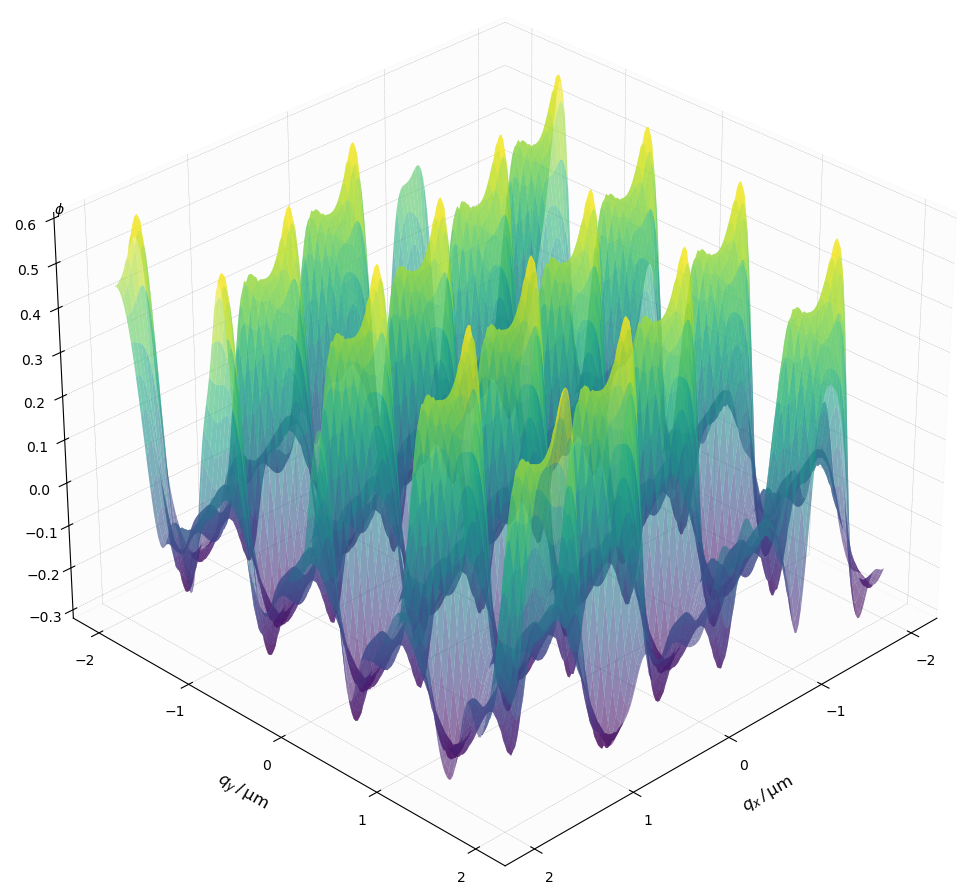}
        \label{3c}
    }
    \subfigure[Di-interstitial (N=61)]{
        \includegraphics[width=0.23\textwidth]{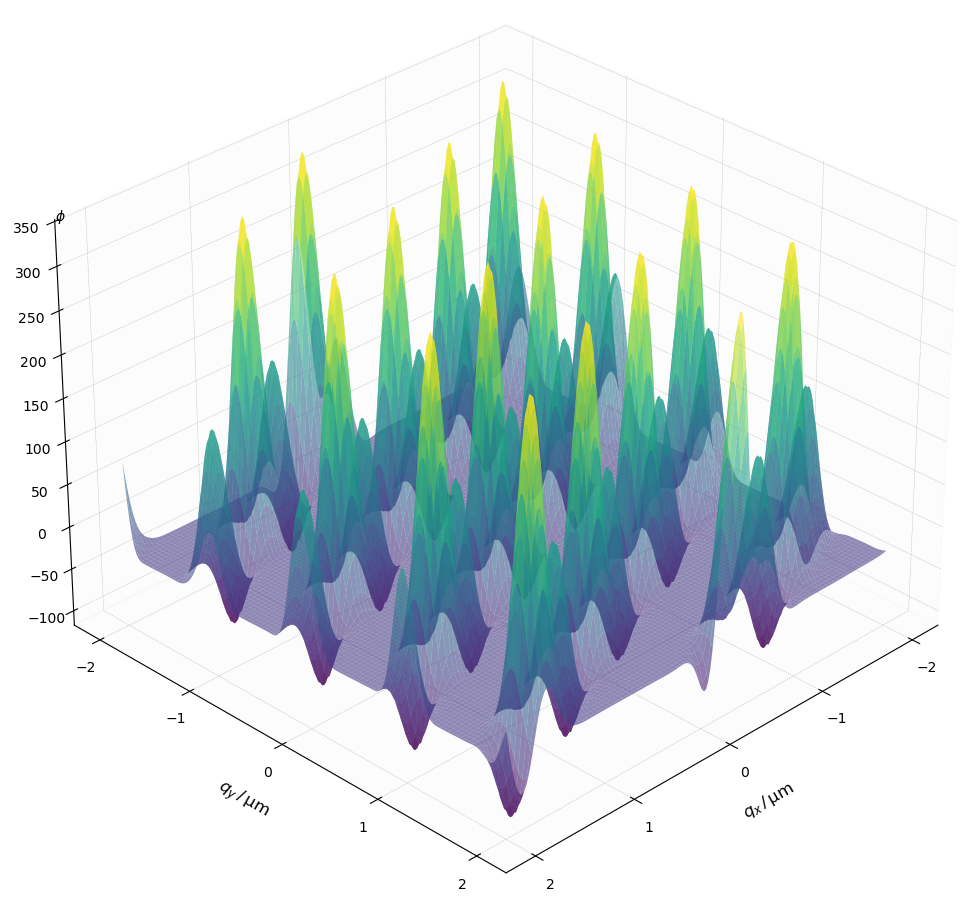}
    \label{3d}
    }
\caption{\label{FIG: Potential}
Periodic stochastic potential landscapes for different defect types. The surface plots visualize the stochastic potential $\phi(\mathbf{q})$ (in units of $k_{\mathrm{B}}T$) obtained under the diffusive approximation.
The landscapes for (a) mono-vacancy, (b) di-vacancy, (c) mono-interstitial, and (d) di-interstitial are shown.
The parameters are the same as in Table \ref{table:diffusion_comparison}, with the number of trajectory points $N$ indicated in parentheses.
}
\end{figure*}

FIG.~\ref{FIG: Potential} presents the stochastic potential $\phi(\mathbf{q})$ derived by combining the diffusive limit [i.e. $\mathbf{Q}=0$ in Eq.~\eqref{dynamical_structure_decomposition}] with the drift and diffusion terms extracted from the trajectories. As $\epsilon$ has been set to $1/2$, the quantity $2\phi(\mathbf{q})$ represents the stochastic energy landscape (in units of $k_\mathrm{B}T$) within the configuration space. It allows us to identify all local minima as well as all saddle points. A quantitative examination of the potential differences between these minima and the range of activation barriers associated with the saddle points follows.

\begin{table*}[t]
\centering
\caption{Comparison of excitation energies and energy barriers. \textbf{Previous studies} (top section) report the free energy difference $\Delta{F}$ (relative to the lowest state) for metastable configurations, estimated from their statistical occurrence~\cite{Pertsinidis_Equilibrium_2001}. \textbf{This work} (bottom section) yields the potential energy difference $\Delta[2\phi]$ for identified local minima and the range of saddle point energies (both relative to the global minimum) from the constructed landscape.
The parameters are the same as in Table \ref{table:diffusion_comparison}.
For comparison, we approximate $\Delta E \equiv \Delta[2\phi] \approx \Delta{F}$, neglecting entropic contributions. The effective number of data points $N$ is in parentheses.}
\label{table:barrier_comparison}
\setlength{\tabcolsep}{0.7em}
\renewcommand{\arraystretch}{1.2}
\begin{tabular}{@{} l c c @{}}
\toprule
\textbf{System and Analysis} & \textbf{Excitation Energy} & \textbf{Energy Barrier Range} \\
& \boldmath $(k_{\mathrm{B}}T)$ & \boldmath $(k_{\mathrm{B}}T)$ \\
\midrule
\textbf{Previous Studies} & & \\
\quad Mono-vacancy (Ref.~\cite{Pertsinidis_Equilibrium_2001}) & 0.24, 0.55 & N/A \\
\quad Di-vacancy (Ref.~\cite{Pertsinidis_Equilibrium_2001}) & 0.4, 1.4 & N/A \\
\cmidrule(r){1-3}
\textbf{Data construction (This Work)} & & \\
\quad Mono-vacancy\footnotemark[1] ($N=600$) & 0.48, 0.57, 0.78, 1.05 & 0.58 -- 1.12 \\
\quad Di-vacancy\footnotemark[2] ($N=600$) & 0.38, 0.61, 0.65, 0.73 & 0.58 -- 0.82 \\
\quad Mono-interstitial\footnotemark[3] ($N=197$) & 0.24 & 0.30 -- 1.41 \\
\quad Di-interstitial\footnotemark[4] ($N=61$) & N/A & N/A \\
\bottomrule
\end{tabular}

\footnotetext[1]{Corresponds to five local minima (lowest and four excited states).}
\footnotetext[2]{Corresponds to five local minima.}
\footnotetext[3]{Corresponds to two local minima.}
\footnotetext[4]{Excluded from quantitative analysis due to insufficient data, leading to physically implausible results (see main text).}
\end{table*}

The local minima of the potential energy landscape correspond to local stable states (potential wells). In a system near equilibrium, the population of a coarse-grained state is proportional to $\exp(-F/k_\mathrm{B}T)$, where $F$ is its free energy. Ling \textit{et al.}~\cite{Pertsinidis_Equilibrium_2001} previously estimated the free energy of excitations for several topologically distinct configurations of mono- and di-vacancies by statistically analyzing their occurrence frequencies over time. We compare these experimentally derived excitation barriers, $\Delta{F}$, with the corresponding $\Delta[2\phi]$, from our inferred landscape (see Table \ref{table:barrier_comparison}), where $\Delta[2\phi]$ is the potential energy difference between a local minimum and the global minimum. For this comparison, we make $\Delta{E}=\Delta(2\phi)\approx\Delta{F}$, effectively neglecting the other minor factors such as the shape of the potential wells.

Despite a limited dataset (e.g., $600$ motion data points for the vacancy case), the computed potential landscape identifies an abundance of features. For both vacancies, we find five distinct local minima. Though the numbers are more than those identified in the finer-grained, configuration-specific examination of Ref.~\cite{Pertsinidis_Equilibrium_2001}, the energy differences between these states are in excellent agreement in order of magnitude ($0.1\text{--}1 \,k_{\mathrm{B}}T$) with the reported $\Delta{F}$ values, demonstrating the effectiveness of our method. It is noteworthy that these agreements are achieved without incorporating any microscopic details of the defect configurations; the potential is obtained solely from the time-series data of the defect positions.

The results for the mono-interstitial ($197$ trajectory points) also yield an energy difference consistent in magnitude with the vacancy cases. In stark contrast, the landscape for the di-interstitial ($61$ points) yields a value that is orders of magnitude larger than those for the other defect types. This severe discrepancy may be attributed to the exceptionally small dataset, which is likely inadequate for a reliable construction. We will therefore exclude the di-interstitial results in subsequent quantitative analysis.

The saddle points in the potential landscape correspond to transition states. The range of activation barriers for state changes may be identified by their lowest and highest saddle points (see Table \ref{table:barrier_comparison}). For both vacancies ($600$ points) and the mono-interstitial ($197$ points), this barrier range is on the order of $0.1$ to $1 \,k_{\mathrm{B}}T$. This energy scale matches that of the free energy differences, consistent with the prior expectation that the activation barriers for the changes would be on the same scale as the free energy differences in this system \cite{Pertsinidis_2005}. Such barriers are readily surmounted at room temperature ($T=295\,\mathrm{K}$), providing a natural explanation for the frequent hops observed experimentally.

This picture also aligns with the diffusion mechanism proposed by Oliveira \textit{et al.}~\cite{Oliveira_mechanism_2011} from simulations of the same system, whereby defect motion occurs through jumps of approximately half a lattice constant concomitant with topological configuration transitions. It differs from mechanisms dominated by the spontaneous generation and annihilation of defects, which require overcoming formation energies---a mechanism observed, for example, in 2D superconducting vortex lattices \cite{Guo_2022}. In the deep solid phase studied here, the formation energy of a point defect is expected to be much larger than the sub-$k_{\mathrm{B}}T$ energy scales we infer, making the generation-annihilation mechanism implausible and favoring the configuration transition mechanism. The concordance between the magnitude of our inferred activation barriers and the energy scale required to explain the frequent configurational changes observed in experiments confirms that the stochastic potential successfully captures the essential system dynamics.

The activation barrier range for the mono-interstitial is broader than for the vacancies. This increased uncertainty is likely attributable to the smaller dataset ($197$ points). Nevertheless, the overall consistency in the order of magnitude supports the general applicability of the method.

\begin{figure*}[t]
    \centering
    \subfigure[Mono-vacancy]{
        \includegraphics[width=0.22\textwidth]{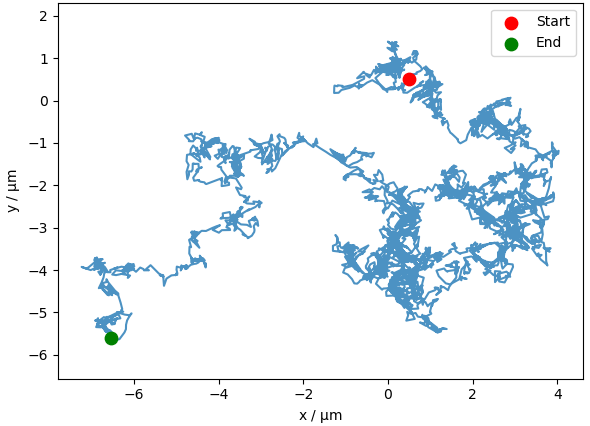}
%        \label{3a}
    }
    \subfigure[Di-vacancy]{
        \includegraphics[width=0.22\textwidth]{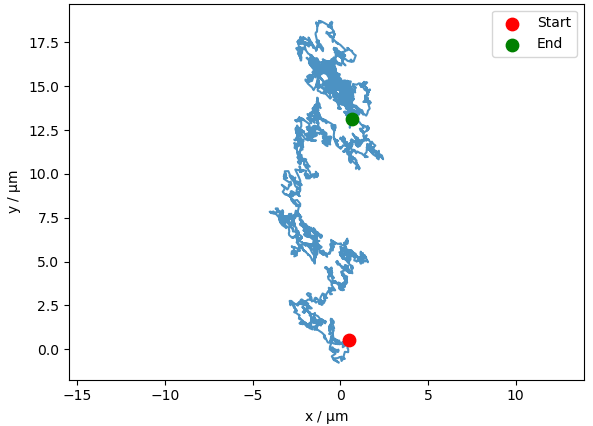}
%    \label{3b}
    }
    \subfigure[Mono-interstitial]{
        \includegraphics[width=0.22\textwidth]{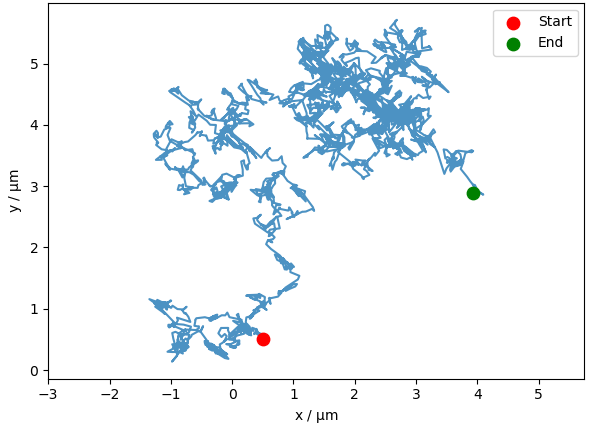}
%        \label{3c}
    }
\caption{\label{Fig: sim_tra}
Stochastic trajectories simulated from the constructed dynamics.
Sample trajectories are generated by numerically integrating the constructed SDEs as described in Sec.~\ref{sec: Sim}. The integration time step is $0.02\ \text{ms}$; for visual clarity, points are plotted at intervals of $2\ \text{ms}$. Panels correspond to (a) mono-vacancy, (b) di-vacancy, and (c) mono-interstitial, with other parameters as in Table \ref{table:diffusion_comparison}.
}
\end{figure*}

\subsection{Qualitative Validation via Simulated Trajectories}

As a final qualitative check, we generated stochastic trajectories within the diffusive limit using the constructed dynamical landscape. Sample trajectories, plotted with a sampling interval of 2 ms, are shown in FIG.~\ref{Fig: sim_tra}.
These paths exhibit a distinct propensity for motion along quasi-linear segments, interspersed with sharp reorientation events---a dynamical signature that closely mirrors experimental observations.
This geometric regularity emerges from the superposition of numerous short-time stochastic fluctuations, reflecting a statistical tendency for defect migration to align preferentially with the principal crystallographic axes over longer time scales.
Furthermore, the characteristic lengths of the straight segments in our simulations are on the order of micrometers, which is consistent with the scales measured experimentally. This alignment suggests that the derived stochastic potential yields energy differences of the correct physical magnitude.

The simulated behavior provides independent, qualitative confirmation that the constructed landscape captures essential dynamical features. The pronounced axial motion, consistent with actual defect trajectories, underscores that the inferred stochastic potential correctly encodes the anisotropic diffusion pathways inherent to the system. Thus, despite approximations arising from limited data, the simulations robustly support the validity of our construction, bridging the statistical model with physical intuition on defect motion.

\section{Discussions}
\label{sec: Discussion}

In this work, we have modeled the stochastic motion of point defects within a 2D hexagonal colloidal crystal. By adopting the framework of evolution mechanics, we represented the defect dynamics via a time-homogeneous SDE [Eq.~\eqref{SDE_both}]. Direct examination of experimental trajectories allowed us to infer the spatially varying drift, $\mathbf{f}(\mathbf{q})$, and diffusion, $\mathbf{D}(\mathbf{q})$, tensors. Crucially, leveraging the periodic boundary conditions inherent to the lattice enabled us to apply the DSD under a diffusive approximation to derive a stochastic potential landscape for each defect type. Finally, by imposing periodicity on the diffusion term, we obtained an SDE suitable for numerical integration via the It\^{o} scheme, enabling the generation of simulated trajectories. The ability to perform a periodic unfolding of the defect coordinate space was foundational to this entire investigation.

Our results yield several significant insights. Most notably, the constructed dynamics reveal that defect motion is distinct from ordinary isotropic Brownian motion. Even in the absence of external driving, the defects experience a non-zero, structured drift field and exhibit an anisotropic, location-dependent diffusion tensor. This characteristic, stemming from the interaction with the underlying periodic lattice, is likely a generic feature of point defect dynamics in other crystalline systems. The stochastic potential landscapes derived from these dynamical terms provide a quantitative energetic picture. The energy differences between identified local minima are in order-of-magnitude agreement with prior experimental estimates, a validation obtained solely from kinematic trajectory data. Furthermore, the estimated range of the activation barriers is consistent with the conjectures of Ling \textit{et al.}~\cite{Pertsinidis_2005}, lending physical credibility to our inferred landscape. It is noteworthy that these coherent results were obtained despite significant data sparsity---a non-trivial fraction of spatial bins contained only a single trajectory point, leading to semi-positive definite diffusion matrices in those positions. This robustness underscores the utility of the evolution mechanics framework for extracting meaningful dynamical and thermodynamic information from limited trajectory data. From a broader methodological perspective, this study demonstrates the power of this framework to dissect complex stochastic dynamics. Finally, our work highlights a clear path for experimental improvement: higher temporal resolution in data acquisition would provide richer local sampling and better reflect temporal continuity, leading to higher-fidelity constructions of the governing SDEs.

Our approach, however, is subject to certain limitations that warrant careful consideration. First, our estimation of the drift and diffusion terms via Eq.~\eqref{discrete_approximations_of_two_terms} inherently assumes the underlying stochastic process obeys the Itô interpretation. The physical applicability of the Itô, Stratonovich, or \textit{A}-type (zero-mass limit) interpretation to the experimental system remains an open question. A comprehensive comparison of results derived under these different stochastic calculi is necessary but non-trivial, and we defer this important investigation to future work. Second, our examination employed the diffusion approximation, neglecting the possible role of a state-dependent $\mathbf{Q}(\mathbf{q})$ matrix. This simplification was pragmatic, avoiding the need for more complex, nonlinear parameterization given the finite dataset. While the final results do not exhibit order-of-magnitude discrepancies, a general system may possess $\mathbf{Q}$-matrix dynamics. Incorporating $\mathbf{Q}(\mathbf{q})$ would require using the more general form of DSD [e.g., Eq.~\eqref{dynamical_structure_decomposition}] for decomposing the drift and would modify the transformation between stochastic interpretations~\cite{shi_relation_2012}.

These limitations naturally point to several promising directions for future research. The most immediate is the acquisition of more extensive trajectory data. With denser sampling, it would become feasible to construct the $\mathbf{Q}(\mathbf{q})$ term and thus obtain a more accurate stochastic potential. A refined landscape would enable the identification of metastable states with greater confidence and yield more reliable thermodynamic (free energy differences) and kinetic (barrier heights) parameters. Furthermore, the geometric structure of such a landscape could elucidate detailed transition pathways between defect states. A reliably inferred potential would also provide a foundation for efficient, large-scale simulations of defect dynamics at the effective-dynamics level.

Beyond this effective-dynamics route, a systematic particle-resolved simulation study would provide a valuable complementary perspective on the microscopic origin and robustness of the periodic landscape. Such a forward-modeling approach would follow a different route from the trajectory-derived effective-dynamics analysis pursued in the present work, in which the simulated trajectories serve as a qualitative consistency check rather than as a particle-resolved model. The two approaches therefore address distinct but complementary levels of description.

Extending the present methodology to more complex defects, such as paired defects which possess internal rotational degrees of freedom and thus inhabit a higher-dimensional state space, presents a rich area for exploration. More generally, the evolution mechanics framework establishes a principled foundation for studying diverse defect types in two-dimensional lattices, applicable to any system where the state can be described by an $n$-dimensional vector $\mathbf{q}(t)$. Applying this framework to systems with higher-dimensional state spaces or more intricate, non-periodic dynamics presents considerable challenges. Fortunately, emerging methodologies that integrate artificial intelligence with time-series analysis~\cite{tang_ying_machine_2025} offer promising avenues to overcome these hurdles, potentially unlocking the study of even more sophisticated stochastic processes in condensed matter and beyond.

\appendix

\renewcommand{\thesection}{}  
\setcounter{equation}{0}      
\renewcommand{\theequation}{A\arabic{equation}}  % A1, A2, ...

\makeatletter
\renewcommand{\@seccntformat}[1]{%
  \ifstrequal{#1}{section}{Appendix: \ }{}
}
\makeatother

\section{Derivation of the Extremal Condition}

This appendix derives the extremal conditions, Eqs.~\eqref{matrix_equation} and \eqref{element_of_A_C}, for the function $W(\{V_{\mathbf{G}}\})$ defined in Eq.~\eqref{SSR}. We aim to express $W$ in a symmetric bilinear form spanning the entire reciprocal space,
\begin{equation}
\label{finally_form}
W = \sum_{\mathbf{G},\mathbf{G}'} A_{\mathbf{G}\mathbf{G}'} V^*_{\mathbf{G}} V_{\mathbf{G}'} + \sum_{\mathbf{G}} \left( C_{\mathbf{G}} V^*_{\mathbf{G}} + C^*_{\mathbf{G}} V_{\mathbf{G}} \right).
\end{equation}
Then we can take $V_{\mathbf{G}}$ and $V^*_{\mathbf{G}}$ as independent variables for variations, which leads to $\partial W / \partial{V^*_{\mathbf{G}}} = 0$ for all $\mathbf{G}$ and yields,
\begin{equation}
\label{extreme_value_condition}
\sum_{\mathbf{G}'} A_{\mathbf{G}\mathbf{G}'} V_{\mathbf{G}'} + C_{\mathbf{G}} = 0.
\end{equation}
This correspends to Eq.~\eqref{matrix_equation} in the main text. We are hence left to identify the coefficients $A_{\mathbf{G}\mathbf{G}'}$ and $C_{\mathbf{G}}$ from Eq.~\eqref{SSR}.

In the residual vector $\mathbf{z}(\mathbf{q}_k)$ at a sampled position $\mathbf{q}_k$, defined in Eq.~\eqref{residual} of the main text, substituting the Fourier series expansion for the potential, Eq.~\eqref{Fourier_series_expansion} gives
\begin{equation}
\label{D_grad_phi_I}
\mathbf{D} \nabla \phi = \sum_{\mathbf{G}} (i\,\mathbf{D}\mathbf{G}) \, V_{\mathbf{G}} \, \exp(i\,\mathbf{G} \cdot \mathbf{q}_k),
\end{equation}
Consider half of the reciprocal space,
\begin{equation}
\label{u}
\mathbf{u} \equiv \sum_{\mathbf{G}}^+ (i\,\mathbf{D}\mathbf{G}) \, V_{\mathbf{G}} \, \exp(i\,\mathbf{G} \cdot \mathbf{q}_k),
\end{equation}
\begin{equation}
\label{u_conj}
\mathbf{u}^* = \sum_{\mathbf{G}}^- (i\,\mathbf{D}\mathbf{G}) \, V_{\mathbf{G}} \, \exp(i\,\mathbf{G} \cdot \mathbf{q}_k),
\end{equation}
where $\sum^+$ ($\sum^-$) denotes summation over one half of the reciprocal space (the complementary half), $V^*_{-\mathbf{G}} = V_{\mathbf{G}}$, and 
$V_0 = 0$, we have
\begin{equation}
\label{D_grad_phi_II}
\mathbf{D} \nabla \phi = \mathbf{u} + \mathbf{u}^*.
\end{equation}
Combining Eqs.~\eqref{residual} and \eqref{D_grad_phi_II} with the definition of $W$ in Eq.~\eqref{SSR} yields an expression in terms of $\mathbf{u}$:
\begin{equation}
\label{SSR_II}
W = \sum_{k} w_k \left[ \mathbf{f}^T\mathbf{f} + 2\mathbf{f}^T(\mathbf{u}+\mathbf{u}^*) + (\mathbf{u}+\mathbf{u}^*)^T (\mathbf{u}+\mathbf{u}^*) \right],
\end{equation}
where the sum $\sum_k$ runs over all sampled positions $\{\mathbf{q}_k\}$ with corresponding weights $w_k$.

We next compute the quadratic and linear terms in $\mathbf{u}$. Using Eqs.~\eqref{u} and \eqref{u_conj}, the quadratic terms in Eq.~\eqref{SSR_II} are:
\begin{align}
\mathbf{u}^{T}\mathbf{u} &= \sum_{\mathbf{G}}^{-}\sum_{\mathbf{G}'}^{+} \left(\mathbf{G}^{T}\mathbf{D}^{T}\mathbf{D}\mathbf{G}'\right) V^*_{\mathbf{G}}V_{\mathbf{G}'} \exp\left[-i \,(\mathbf{G}-\mathbf{G}') \cdot \mathbf{q}_k\right], \label{uT_u} \\
\mathbf{u}^{*T}\mathbf{u}^* &= \sum_{\mathbf{G}}^{+}\sum_{\mathbf{G}'}^{-} \left(\mathbf{G}^{T}\mathbf{D}^{T}\mathbf{D}\mathbf{G}'\right) V^*_{\mathbf{G}}V_{\mathbf{G}'} \exp\left[-i \,(\mathbf{G}-\mathbf{G}') \cdot \mathbf{q}_k\right], \label{u*T_u*} \\
\mathbf{u}^{*T}\mathbf{u} &= \sum_{\mathbf{G}}^{+}\sum_{\mathbf{G}'}^{+} \left(\mathbf{G}^{T}\mathbf{D}^{T}\mathbf{D}\mathbf{G}'\right) V^*_{\mathbf{G}}V_{\mathbf{G}'} \exp\left[-i \,(\mathbf{G}-\mathbf{G}') \cdot \mathbf{q}_k\right] \nonumber \\
&= \sum_{\mathbf{G}}^{-}\sum_{\mathbf{G}'}^{-} \left(\mathbf{G}^{T}\mathbf{D}^{T}\mathbf{D}\mathbf{G}'\right) V^*_{\mathbf{G}}V_{\mathbf{G}'} \exp\left[-i \,(\mathbf{G}-\mathbf{G}') \cdot \mathbf{q}_k\right]. \label{u*T_u}
\end{align}
Combining Eqs.~\eqref{uT_u}--\eqref{u*T_u} gives the complete quadratic form:
\begin{align}
\left( \mathbf{u}+\mathbf{u}^* \right)^T \left( \mathbf{u}+\mathbf{u}^* \right) =& \,\mathbf{u}^{T}\mathbf{u} + \mathbf{u}^{*T}\mathbf{u}^* + 2\mathbf{u}^{*T}\mathbf{u} \nonumber \\
=& \sum_{\mathbf{G}}\sum_{\mathbf{G}'} \left(\mathbf{G}^{T}\mathbf{D}^{T}\mathbf{D}\mathbf{G}'\right) V^*_{\mathbf{G}}V_{\mathbf{G}'} \exp\left[-i \,(\mathbf{G}-\mathbf{G}') \cdot \mathbf{q}_k\right]. \label{quadratic_terms}
\end{align}
For the linear term, using the fact that $\mathbf{u} + \mathbf{u}^* = \mathbf{D}\nabla\phi$ [see Eqs.~\eqref{D_grad_phi_I} and \eqref{D_grad_phi_II}] is real, we find
\begin{align}
2(\mathbf{u} + \mathbf{u}^*) =&  \,\mathbf{D}\nabla\phi + (\mathbf{D}\nabla\phi)^*
\nonumber \\
=& \sum_{\mathbf{G}} [ (i\,\mathbf{D}\mathbf{G}) V_{\mathbf{G}} \exp(i\,\mathbf{G}\cdot \mathbf{q}_k)  \nonumber \\
&+ (-i\,\mathbf{D}\mathbf{G}) V^*_{\mathbf{G}} \exp(-i\,\mathbf{G}\cdot \mathbf{q}_k) ]. \label{linear_term}
\end{align}
Finally, substituting Eqs.~\eqref{quadratic_terms} and \eqref{linear_term} into Eq.~\eqref{SSR_II} and rearranging, we obtain the desired form:
\begin{align}
W =& \sum_{k} w_k\, \mathbf{f}^T\mathbf{f} + \sum_{\mathbf{G}} \left[ V_{\mathbf{G}} \sum_{k} w_k \left( i\,\mathbf{f}^T\mathbf{D}\mathbf{G} \right) \exp(i\,\mathbf{G}\cdot \mathbf{q}_k) \right. \nonumber \\
&+
\left. V_{\mathbf{G}}^* \sum_{k} w_k \left(-i\,\mathbf{f}^T\mathbf{D}\mathbf{G} \right) \exp(-i\,\mathbf{G}\cdot \mathbf{q}_k) \right] \nonumber \\
&+ \sum_{\mathbf{G}}\sum_{\mathbf{G}'} V^*_{\mathbf{G}}V_{\mathbf{G}'} \sum_{k} w_k \left(\mathbf{G}^{T}\mathbf{D}^{T}\mathbf{D}\mathbf{G}'\right) \exp\left[-i \,(\mathbf{G}-\mathbf{G}') \cdot \mathbf{q}_k\right] \nonumber \\
=& \sum_{k} w_k\, \mathbf{f}^T\mathbf{f} + \sum_{\mathbf{G}} \left[ C_{\mathbf{G}}^* V_{\mathbf{G}} +
C_{\mathbf{G}} V_{\mathbf{G}}^* \right] \nonumber \\
&+ \sum_{\mathbf{G}}\sum_{\mathbf{G}'} A_{\mathbf{G}\mathbf{G}'} V^*_{\mathbf{G}}V_{\mathbf{G}'}, \label{SSR_III}
\end{align}
Comparing Eq.~\eqref{SSR_III} with Eq.~\eqref{finally_form} yields the coefficients $A_{\mathbf{G}\mathbf{G}'}$ and $C_{\mathbf{G}}$, as stated in Eq.~\eqref{element_of_A_C} of the main text.

\begin{acknowledgments}
Xicheng Huang would like to thank Professor Xiaomei Zhu for her guidance and Binglin Zhu for stimulating discussions.
This work was supported in part by the National Natural Science Foundation of China (YCC, Approval No. 12375034) and by the Sichuan University Talent Introduction Fund.
\end{acknowledgments}
%\nolinenumbers
\bibliography{MyBiblio}
\end{document}